\documentclass{aa}  
\usepackage{graphicx}
\usepackage{txfonts}
\newcommand{\mygi}{MyGIsFOS}
\newcommand{\vr}{\ensuremath{V_\mathrm{r}}}

\newcommand{\logg}{\ensuremath{\log\,g}}
\def\teff{$T\rm_{eff}$}
\newcommand{\kms}{$\rm km s ^{-1}$}

\begin{document} 

\title{The Gaia RVS benchmark stars\\ 
II. A sample of stars selected for their Gaia high radial velocity
\thanks{Based on observations made with UVES at VLT 0109.22XP.001 and 110.23V0.001.}
}
\titlerunning{Gaia RVS benchmark stars II}

\author{
E.~Caffau    \inst{1} \and
D.~Katz \inst{1} \and          
A. G\'omez \inst{1} \and
P.~Bonifacio \inst{1} \and
R.~Lallement \inst{1} \and      
P.~Sartoretti \inst{1} \and    
L.~Sbordone \inst{2} \and      
M.~Spite \inst{1} \and
A.~Mucciarelli \inst{3} \and
R.~Ibata \inst{4} \and
L.~Chemin \inst{5} \and        
F.~Th\'evenin \inst{6} \and  
P.~Panuzzo \inst{1} \and      
N.~Leclerc \inst{1} \and    
P.~Fran\c{c}ois \inst{7,8} \and
H.-G.~Ludwig \inst{9} \and    
L.~Monaco \inst{10} \and
M.~Haywood \inst{1} \and
C.~Soubiran \inst{11}
}

\institute{GEPI, Observatoire de Paris, Universit\'{e} PSL, CNRS,  5 Place Jules Janssen, 92190 Meudon, France
\and
European Southern Observatory, Casilla 19001, Santiago, Chile
\and
Dipartimento di Fisica e Astronomia, Universit\`a degli Studi di Bologna, Via Gobetti 93/2, I-40129 Bologna, Italy
\and
Universit{\'e} de Strasbourg, CNRS, Observatoire astronomique de Strasbourg, UMR 7550, F-67000 Strasbourg, France
\and
Instituto de Astrof\'isica, Departamento de Ciencias F\'isicas, Universidad Andr\'es Bello, Fernandez Concha 700, Las Condes, Santiago, Chile
\and
Universit\'e de Nice Sophia-Antipolis, CNRS, Observatoire de la C\^ote d'Azur, Laboratoire Lagrange, BP 4229, F-06304 Nice, France
\and
GEPI, Observatoire de Paris, Universit\'{e} PSL, CNRS, 77 Av. Dendert-Rocheau, 75014 Paris, France
\and
UPJV, Universit\'e de Picardie Jules Verne, 33 rue St Leu, 80080 Amiens, France
\and
Zentrum f\"ur Astronomie der Universit\"at Heidelberg, Landessternwarte, K\"onigstuhl 12, 69117 Heidelberg, Germany
\and
Instituto de Astrof{\'\i}sica, Departamento de Ciencias F{\'\i}sicas, Universidad Andres Bello, 
Autopista Concepci{\'o}n-Talcahuano, 7100, Chile
\and
Laboratoire d’Astrophysique de Bordeaux, Univ. Bordeaux, CNRS, B18N, all{\'e}e Geoffroy Saint-Hilaire, 33615 Pessac, France
}

   \date{Received September 15, 1996; accepted March 16, 1997}

  \abstract
{The Gaia satellite has already provided the astronomical community with three data releases, and the Radial Velocity Spectrometer (RVS) 
on board Gaia has provided the radial velocity for 33 million stars.
}
{When deriving the radial velocity from the RVS spectra, several stars are measured to have large values.
To verify the credibility of these measurements, we selected some bright stars with the modulus of
radial velocity
in excess of 500\,\kms\ to be observed with SOPHIE at OHP and UVES at VLT.
This paper is devoted to investigating the chemical composition of the stars observed with UVES.
}
{We derived atmospheric parameters using Gaia photometry and parallaxes, and we
performed a chemical analysis using the \mygi\ code.
}
{We find that the sample consists of metal-poor stars, although none have extremely low metallicities.
The abundance patterns match what has been found in other samples of metal-poor stars selected
irrespective of their radial velocities. We highlight the presence of three stars with low Cu and Zn
abundances that are likely descendants of pair-instability supernovae. Two stars are apparently
younger than 1\,Ga, and their masses exceed twice the turn-off mass of metal-poor populations.
This makes it unlikely that they are blue stragglers because it would imply they
formed from triple or multiple systems. We suggest instead that they are young
metal-poor stars accreted from a dwarf galaxy. Finally, we find that the star RVS721 is associated with
the Gjoll stream, which itself is associated with the Globular Cluster NGC\,3201.}
{}

   \maketitle
%
\section{Introduction\label{intro}}
  
Since 2014 the Gaia satellite \citep{gaiacol16} has provided crucial information (most important positions, parallaxes, proper motions, radial velocities, and
stellar parameters) for a huge sample of stars.
Since its launch, there have been three data releases \citep[DR1, DR2, DR3;][]{gaiadr1,gaiadr2,gaiadr3}.

The data provided in the Gaia releases have been used by the astronomical community to improve our knowledge
on the Milky Way and the Local Group galaxies.
There are still limitations on the data provided by Gaia, such as a limitation in brightness,
and in this respect, there is nothing that can be done to overcome it.
But there are things that can be improved.
In \citet{rvsBM1}, we presented a project to build an archive of Gaia Radial Velocity Spectrometer (RVS) benchmark stars, acquiring high-quality spectra to be used to derive the
line spread function of stars observed by the Gaia RVS and to provide reference stars for the radial velocity (\vr) determination
in as many focal plane positions as possible.
In this way, we had the goal of improving the \vr\ determination from the RVS spectra.
The project is still open, and the spectra observed will be used in the Gaia data releases 4 and 5.
Nine of the stars investigated in this work (RVS701, RVS702, RVS703, RVS706, RVS707, RVS714, RVS719, RVS720) have spectra of sufficient quality
in the RVS wavelength range and will be added into the archive of Gaia RVS benchmark stars.

This project also gives us the opportunity to investigate
other possible problems for the \vr\ measurements from the RVS spectra.
The RVS spectra have a wavelength range of less than 30\,nm, and several spectra have a low signal-to-noise ratio (S/N);
in fact, a considerable fraction (about 25\%) of \vr\ determination in the Gaia\,DR3 derives from spectra with $\rm S/N < 5$ \citep[see][]{katz23}.
These facts, especially the latter, introduce a limitation for a good \vr\ determination.
We also found in the Gaia DR\,3 that there are stars with extreme \vr, and it is unclear whether  
these values are real or just evidence of problematic spectra.
To verify the extreme \vr\ values and confirm or discard them, we added a sample of extreme \vr\ stars
in the project focused on observing the Gaia RVS benchmark stars.
The kinematic investigation of the sample is to be presented by Katz et al. (in preparation). In this work, we show the results of 
our chemical investigation.

Another intriguing fact of this high \vr\ sample is that these stars are surely high-speed stars,
and it is indeed interesting to combine kinematic and chemical characteristics of high-speed stars.
\citet{ghs1} investigated a sample of stars selected in Gaia\,DR\,2 for their high transverse velocity,
so still high-speed stars. In their sample, some stars happen to appear younger than 6 Ga.
This is an intriguing finding that could be explained by a population of blue straggler stars.
From a similar selection from Gaia\,DR3, \citet{bonifacio23} highlighted a metal-poor, apparently young population
barely compatible with blue straggler stars because their masses are in excess of twice the mass of a typical turn-off metal-poor star
and in excess of typical masses of blue stragglers in galactic globular clusters and in the field.
We questioned whether we would also find this apparently young population in high \vr\ stars.
The answer we derived is positive, and the two young stars detected in this work are unlikely to be blue stragglers because 
their masses are in excess of $\rm 1.9 M_\odot$, while blue straggler stars typically have masses  $\rm  < 1.2 \ to\ 1.5  M_\odot$
\citep[see e.g.][]{carney05,fiorentino14,ferraro23}.

\section{Target selection\label{sel}} 

The targets we selected from the Gaia\,EDR3 \citep{GaiaEDR3} and Gaia\,DR3 \citep{gaiadr3} catalogues 
are stars with an absolute radial velocity in excess of 500\,\kms.
Overall, we wanted to verify whether these large values are mistakes of the pipeline or that the stars really have a large radial velocity.

We made a similar selection of targets regarding a sample of stars observed with SOPHIE at Observatoire de Haute-Provence (OHP).
The SOPHIE sample and the kinematic analysis of this sample are discussed in Katz et al. (in preparation).
As stated in Katz et al. (in preparation), the stars observed with UVES and investigated in this work are high \vr\ stars.
The Gaia G magnitude in the sample is in the range from 10.2 to 13.6\,mag.

\section{Observations} 

A sample of 45 stars 
was observed in the ESO programme 0109.22XP.001 with UVES \citep{uves} at VLT in the setting DIC2\,437+760
(wavelength ranges 373-499 and 565-946\,nm),
with slit 0\farcs{4}\ (resolving power 90\,000) in the  blue arm and 0\farcs{3}\ (resolving power 110\,000) in the red arm.
A single exposure of about half an hour was taken for each star.
A sub-sample of 19 stars were re-observed in ESO programme 110.23V0.001.

Of the observations in P109, 42 were graded 'A' and three 'B'.
In P110, of the 21 spectra investigated, 17 were graded 'A' and four 'B'.
For the star RVS735, observed in P109, there is no blue spectrum.

\begin{figure*}
\centering
\includegraphics[width=\hsize,clip=true]{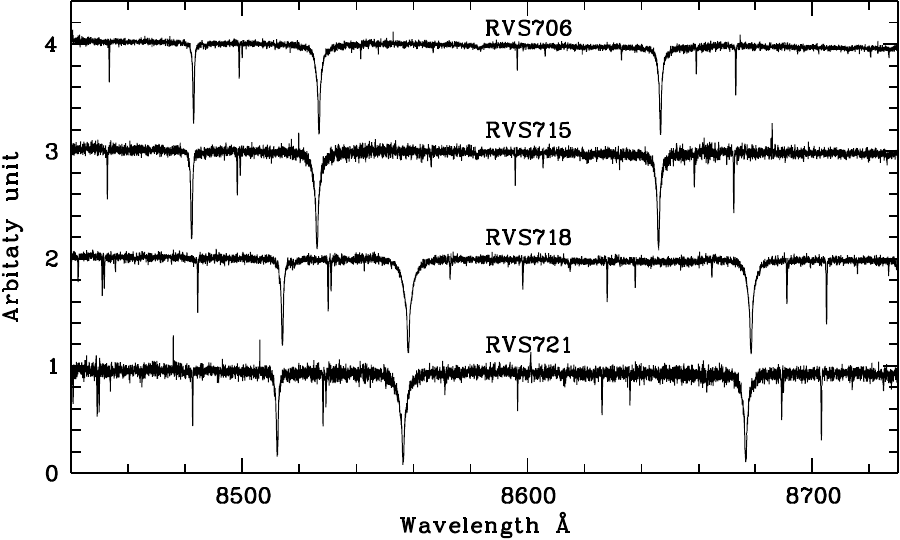}
\caption{Four spectra shown in the wavelength range of the Gaia RVS.}
\label{fig:obs}
\end{figure*}

For the stars of P109, we downloaded the UVES spectra reduced by ESO in order to have 
an homogeneous data reduction and to make data readily available to the community.
The spectra of four stars are shown in Fig.\,\ref{fig:obs} in the wavelength range of Gaia RVS.
Few stars showed problems in the data reduction, and we reduced the spectra by using the ESO pipeline.
For P110, we reduced the spectra to have quick access to them.

\begin{figure*}
\centering
\includegraphics[width=\hsize,clip=true]{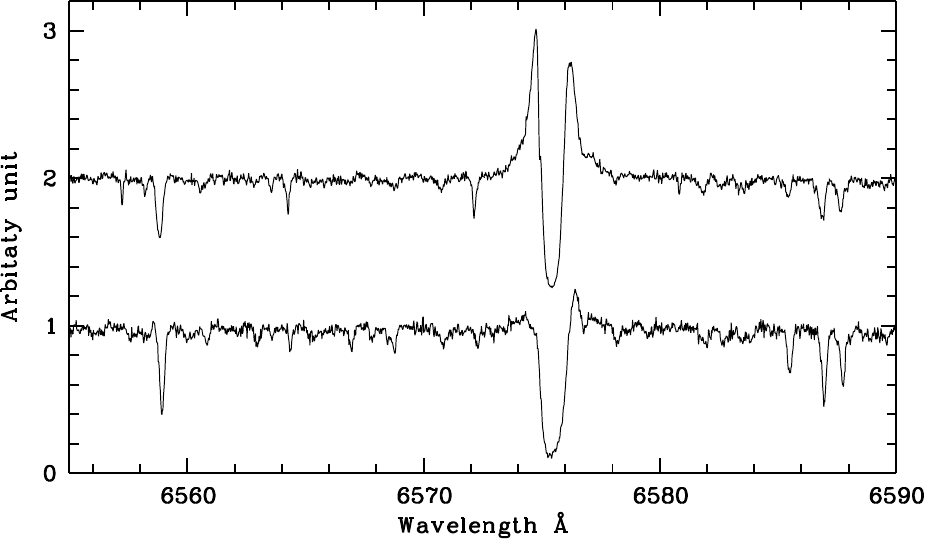}
\caption{Two observations of RVS724.}
\label{fig:obsrvs724}
\end{figure*}

We performed the chemical analysis of 43 stars.
Regarding the two stars for which there is no chemical analysis, 
the spectrum of RVS705 is noisy (S/N is too low) and crowded due to it being a carbon star \citep{alksnis01}. The second star,
RVS724, is a variable and active star whose two spectra are different (see Fig.\,\ref{fig:obsrvs724}). 
For the stars with observations in P109 and P110, we added the spectra, except for RVS712 and RVS736, which showed
clear differences in radial velocity.

\section{Analysis}

The kinematic investigation of this sample of stars is presented in Katz et al. (in preparation).
The stars investigated in this work all belong to the Galactic halo according to the criteria by \citet{bensby14}.

For the chemical investigation, we needed the spectrum to be at laboratory wavelength.
We applied the radial velocity provided by Gaia\,DR3.
All the stars we investigated have been confirmed to have a high radial velocity.
For the majority of them, the velocity we derived is within a few kilometres per second from the value provided by Gaia\,DR3.
Just one star, RVS739, shows a disagreement between the two velocities, by about 10\,\kms, but the star is still a high \vr\ star.
Our pipeline \mygi\ for the chemical investigation was allowed to shift each feature by half of the total broadening
of the grid used in the analysis of each star. We verified that the shift 
applied by \mygi\ was sufficient to investigate all the features. If it was not, we applied an extra shift to the spectrum.

\subsection{Stellar parameters} \label{sec:param}

To derive the stellar parameters, we used the Gaia\,DR3 photometry and parallax and the reddening from 
\citet{vergely22}. 
We derived the parallax zero-point as suggested by \citet{lindegren21}.
We first de-reddened the Gaia $G, G_{BP}$, and $G_{RP}$ photometry by using the maps provided by \citet{vergely22} 
and the theoretical extinction coefficients derived from model atmospheres (Mucciarelli \& Bonifacio in preparation).
The $G_{BP}-G_{RP}$ Gaia\,DR3 colours were then compared to the synthetic photometry in order to derive the effective temperature.
The gravity was derived by using the parallax, corrected by the zero-point, through the Stefan-Boltzmann equation.
We first assumed a solar metallicity and derived for each star the effective temperature, \teff, and surface gravity, \logg.

The initial stellar parameters were used to derive the metallicity of the stars by using \mygi\ \citep{mygi14}.
The metallicity derived in this way was then used to derive new stellar parameters.
The iterations were repeated up to when the changes in \teff\ and \logg\ were within 10\,K and 0.02\,dex, respectively.
The micro-turbulence ($\xi$) was derived by using the calibration of \citet{mashonkina17}. 
For the majority of the stars in the sample, the signal-to-noise ratio of the spectra 
is in fact too low to allow the use of lines with small equivalent widths 
in order to derive the micro-turbulence by the balance of A(Fe) derived from \ion{Fe}{i} lines of different equivalent widths.

Regarding the derivation of the stellar parameters, the grid of synthetic photometry was limited, and we did not allow any extrapolation in \teff\ or \logg\ beyond the grid.
The lower limit accepted in \teff\ is 4000\,K.
Six stars converge at this lower-limit in \teff\ when deriving the stellar parameters, so one could suppose that these stars are in fact cooler.
For all these stars (RVS708, RVS714, RVS719, RVS727, RVS730, and RVS734) the calibration of \citet{mucciarelli21} provides values close to 4000\,K,
so we expected that by adopting 4000\,K, the stellar \teff\ would be well within the uncertainty.

If we had used the distance estimate of  \citet{2021AJ....161..147B} provided in Gaia\,DR3 to derive the stellar parameters instead of the parallax corrected
by the zero-point, we would have derived \teff\ as being, on average, 6\,K hotter (with a difference always within 40\,K) 
and an average difference in \logg\ of --0.07\,dex (with a maximum value at --0.21\,dex).
These changes in the stellar parameters would not have affected the metallicity derived from \ion{Fe}{i} lines
(generally well within 0.1\,dex and 0.02\,dex for \teff\ and \logg, respectively) and would have only marginally affected 
(within 0.1\,dex) the iron ionisation balance because the Fe abundance, when derived from \ion{Fe}{ii}, is sensitive to the surface gravity.

When comparing our adopted stellar parameters (\teff\ and \logg) to the parameters derived by using the calibration of \citet{mucciarelli21},
we found an average difference in \teff\ of $-73\pm 37$\,K\footnote{Our values minus \citet{mucciarelli21} values.} 
and in \logg\ of $-0.03\pm 0.02$\,dex.
We adopted 100\,K as the uncertainty in \teff\ and 0.1\,dex in \logg.

\begin{figure}
\centering
\includegraphics[width=\hsize,clip=true]{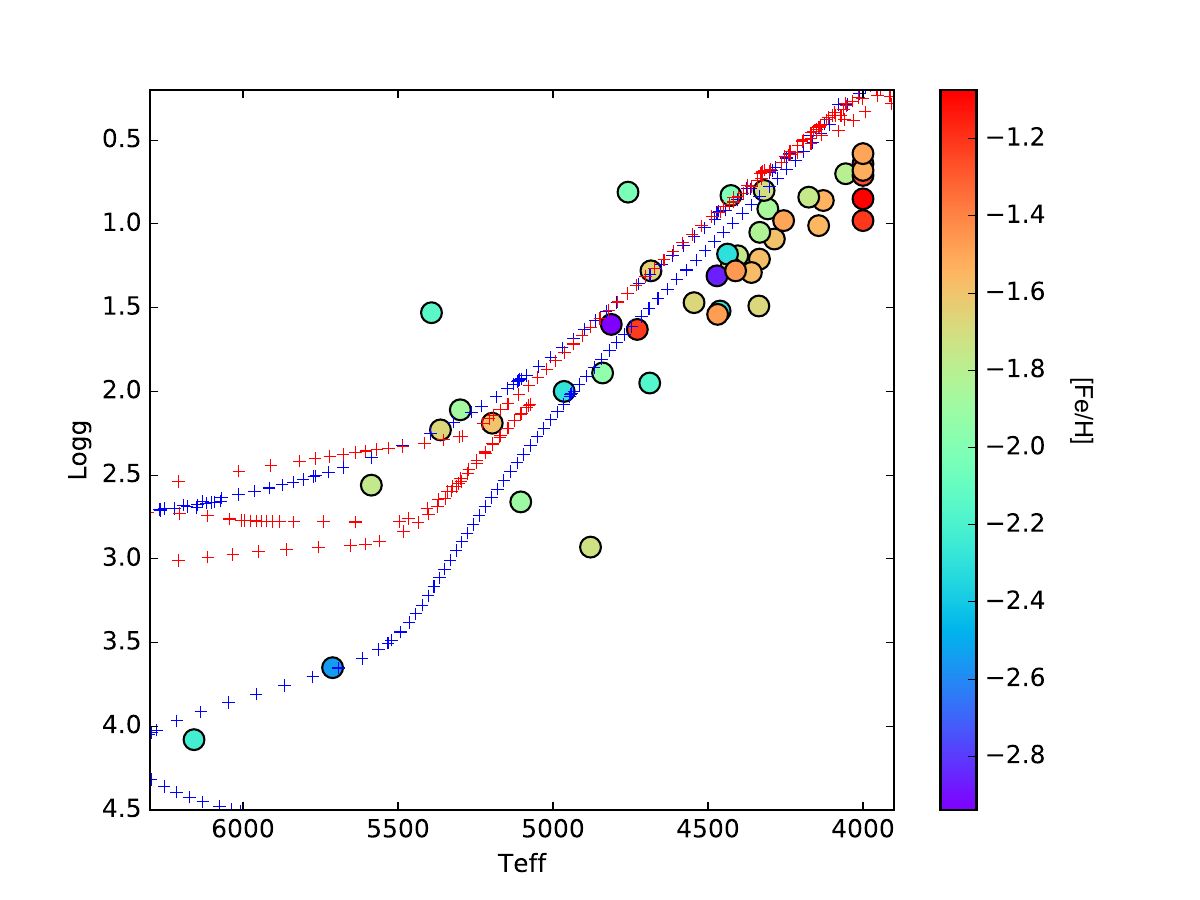}
\caption{Observed stars in the (\teff,\logg) diagram compared to PARSEC isochrones
of metallicity --1.5\,dex and of ages 12\,Ga (blue crosses) and 1\,Ga (red crosses), to guide the eye.
The [Fe/H] of the stars are colour coded according the vertical bar on the left.
}
\label{fig:plotiso}
\end{figure}

In Fig.\,\ref{fig:plotiso}, a comparison of the adopted stellar parameters to PARSEC isochrones \citep{bressan12,marigo17} is shown.
In the figure we noted that at least two stars (RVS725 and RVS726), which are both very metal poor, 
appear younger than expected when compared to the isochrones.
When comparing the $G_{BP}-G_{RP}$ colour and the $G$ absolute magnitude to isochrones of various ages,
depending on the evolutionary stage, the possible expected masses and ages for RVS725 are 
$\rm 3.2 M_\odot$ with age of 230\,Ma (as a sub-giant branch or a Hertzsprung gap for more intermediate and massive stars), 
$\rm 2.9 M_\odot$ with age of 280\,Ma (still core He-burning, the blueward part of the Cepheid loop of intermediate and massive stars), 
$\rm 2.6 M_\odot$ with age of 390\,Ma (the early asymptotic giant branch or a quick stage of red giant for massive stars\footnote{\url{http://stev.oapd.inaf.it/cmd_3.1/faq.html}}). 
For RVS726 we have just one solution in the interpolation, which is
$\rm 1.9 M_\odot$ with age of 900\,Ma (the early asymptotic giant branch or a quick stage of red giant for massive stars). 
A mass below about $\rm 1.8 M_\odot$ can in principle be acceptable for a star to be a blue straggler that is the result 
of the merging of two old metal-poor stars. However, one has to assume no mass was lost in the merging process.
The mass expected for RVS725 is far too large.
For both stars, to derive the masses and ages, we used a PARSEC isochrone of metallicity of $-1.95$
(we emphasise that PARSEC isochrones are not enhanced in $\alpha$ elements, and for this reason, we selected a metallicity that was slightly higher
than the [Fe/H] derived for the two stars, which are $\alpha$-enhanced) and ages from 100\,Ma and 3\,Ga with a step of 200\,Ma.
An isochrone at metallicity $-1.65$\,dex would provide the same result or increase the masses by less than $\rm 0.1 M_\odot$.
We concluded that these two stars, but especially RVS725, can really be young stars and were probably formed in dwarf galaxies in a recent time
and then accreted.

\begin{table}
\caption{Solar abundances.}
\label{tab:solarabbo}
\begin{tabular}{lll}
\hline
Element & A(X) & Reference \\
\hline
C  & 8.50 & \citet{abbosunEC} \\
O  & 8.76 & \citet{abbosunEC} \\
Na & 6.30 & \citet{lodders09} \\
Mg & 7.54 & \citet{lodders09} \\
Al & 6.47 & \citet{lodders09} \\
Si & 7.52 & \citet{lodders09} \\
S  & 7.16 & \citet{abbosunEC} \\
Ca & 6.33 & \citet{lodders09} \\
Sc & 3.10 & \citet{lodders09} \\
Ti & 4.90 & \citet{lodders09} \\
V  & 4.00 & \citet{lodders09} \\
Cr & 5.64 & \citet{lodders09} \\
Mn & 5.37 & \citet{lodders09} \\
Fe & 7.52 & \citet{abbosunEC} \\
Co & 4.92 & \citet{lodders09} \\
Ni & 6.23 & \citet{lodders09} \\
Cu & 4.21 & \citet{lodders09} \\
Zn & 4.62 & \citet{lodders09} \\
Y  & 2.21 & \citet{lodders09} \\
Eu & 0.52 & \citet{lodders09} \\
Ba & 2.17 & \citet{lodders09} \\
\hline
\end{tabular}
\end{table}

The stars analysed in this work are all metal poor ($\rm [Fe/H]<-1.0$), but no star is extremely metal poor ($\rm [Fe/H]<-3$).
We recall that RVS705 is metal rich, but this star has not been analysed.
The stellar parameters and the abundances derived are provided at CDS in an online table that is described in the appendix.

\begin{table}
\caption{Signal-to-noise ratio of the spectra.}
\label{tab:snr}
\begin{tabular}{llll}
\hline
Star & ID Gaia\,DR3 & Period & S/N  \\
     &              &        &     @600\,nm \\
\hline
RVS700 & 3064774107059078784 & 109     & 75 \\
RVS701 & 6148919860947868160 & 109+110 & 78 \\
RVS702 & 5654083549060480256 & 109     & 66 \\
RVS703 & 4585522112754440576 & 109     & 82 \\
RVS704 & 4567509294790548096 & 109     & 68 \\
RVS705 & 5641794449335693184 & 109     & $<$20 \\
RVS706 & 1772048878641647488 & 109+110 & 76 \\
RVS707 & 6102498720545983744 & 109+110 & 65 \\
RVS708 & 4087001612275854976 & 109     & 38 \\
RVS709 & 1774647123402000896 & 109     & 32 \\
RVS710 & 2831494371421184512 & 109     & 24 \\
RVS711 & 5806001524479916160 & 109+110 & 37 \\
RVS712 & 5843949763886962304 & 109     & 34 \\
RVS712 & 5843949763886962304 & 110     & 49 \\
RVS713 & 2731609959149919360 & 109     & 40 \\
RVS714 & 1789332097623284736 & 109     & 45 \\
RVS715 & 4200211044628918016 & 109     & 23 \\
RVS716 & 3545088275525629312 & 109+110 & 42 \\
RVS718 & 5430581735975161344 & 109+110 & 44 \\
RVS719 & 4048087421869545216 & 109+110 & 46 \\
RVS720 & 4292051361144994304 & 109     & 27 \\
RVS721 & 5365576065922008960 & 109+110 & 26 \\
RVS722 & 5932387366845310208 & 109     & 32 \\
RVS723 & 3830347565099377152 & 109+110 & 67 \\
RVS724 & 5354656541072512000 & 109     &    \\
RVS725 & 5860126260025908992 & 109     &  8 \\
RVS726 & 5515555464906936192 & 109     & 15 \\
RVS727 & 4052560510049395840 & 109+110 & 40 \\
RVS728 & 3585461964540660224 & 109+110 & 36 \\
RVS729 & 4434477664957276544 & 109+110 & 26 \\
RVS730 & 1761871661577694848 & 109     & 26 \\
RVS731 & 1751668601691825664 & 109     & 19 \\
RVS732 & 4532552345513094144 & 109     & 10 \\
RVS733 & 3572878053960750976 & 109+110 & 51 \\
RVS734 & 4207775547189676160 & 109+110 & 28 \\
RVS735 & 2697864023148222976 & 109     & 14 \\
RVS736 & 5806698825316687872 & 109     & 22 \\
RVS736 & 5806698825316687872 & 110     & 18 \\
RVS737 & 6418433113222352000 & 109+110 & 26 \\
RVS738 & 1876742967089592448 & 109     & 20 \\
RVS739 & 1775026115611202560 & 109     & 21 \\
RVS740 & 6103440486613002496 & 109+110 & 51 \\
RVS741 & 4219344020113924096 & 109     & 17 \\
RVS742 & 1803730786507192448 & 109     & 24 \\
RVS743 & 4638362923593318912 & 109     & 29 \\
RVS744 & 2689355276322100224 & 109     & 18 \\
RVS745 & 5803679463301298944 & 109+110 & 25 \\
\hline
\end{tabular}
\end{table}

\subsection{Abundances}\label{secabbo}

The abundances were derived by \mygi\ \citep{mygi14}.
The solar reference abundances are listed in Table\,\ref{tab:solarabbo}.
The synthetic spectra grids are the same as described in 
\citet{rvsBM1}.

Lithium is clearly visible in the metal-poor turn-off star RVS738, providing $\rm A(Li)_{\rm LTE}=2.34$.
Three evolved stars (RVS702, RVS706, RVS711) show a feature at the wavelength position of the Li 670\,nm doublet,
and the Li abundance derived from this feature is low ($\rm A(Li)_{\rm LTE}$ of --0.30, 1.07 and 0.89, respectively).

For some elements, we compared the observed spectrum to a grid of synthetic spectra, 
computed with SYNTHE from an ATLAS\,9 model \citep{K05} while only varying the abundance of one element. 
We minimised the $\chi^2$ to derive the abundance.

\begin{figure}
\centering
\includegraphics[width=\hsize,clip=true]{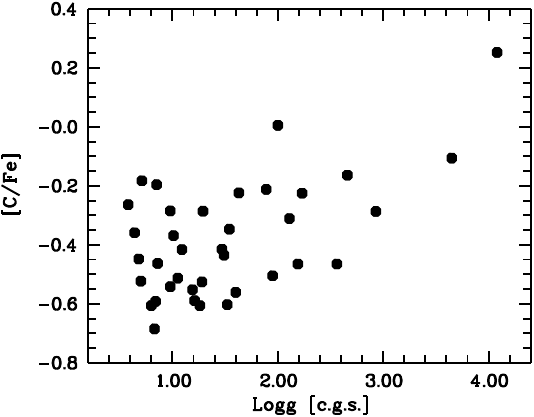}
\caption{[C/Fe] versus \logg\ for the observed sample.
}
\label{fig:loggcfe}
\end{figure}

We investigated the G-band at 430\,nm formed by CH lines to verify if any of the stars was a carbon-enhanced metal-poor (CEMP) star.
Through line profile fitting, we could derive the C abundance for 39 stars in the sample, and none appeared to be rich in carbon.
We found nearly all stars but one have a negative or close to zero [C/Fe] ratio, in agreement with
the results by \citet{FS6}.
These stars are all evolved, and the carbon destruction is due to their stellar evolution.
The turn-off star RVS738, as expected and described in \citet{FS12}, shows a C enhancement: $\rm [C/Fe]=+0.25\pm 0.25$.
In Fig.\,\ref{fig:loggcfe}, the [C/Fe] ratio is plotted against the surface gravity. We expected that more evolved stars would have 
converted more C into N, but things are more complicated than that. The trend is not evident, and an extra-mixing
should be present, though different in the various stars, in order to favour C destruction.
We also expected at least a fraction of the stars to be CEMP stars, but
we found none.
In this metallicity regime \citep{topos2}, we expected that CEMP stars would lie 
on the high C band described by \citet{spite13}, with a carbon abundance just below solar. 
So [C/Fe] would still be positive even if the evolution of the CEMP stars had destroyed part of their carbon.

Oxygen was derived by line profile fitting because the region is contaminated by telluric absorptions.
We derived A(O) from the [OI] lines for 26 stars in the sample.
All the stars in the sample are enhanced in oxygen, with $\rm \langle [O/Fe]\rangle =0.60\pm 0.11$, which is consistent with the results by \citet{FS6},
and the star with the highest oxygen value was RVS733, $\rm [O/Fe] =0.92$.

\begin{figure}
\centering
\includegraphics[width=\hsize,clip=true]{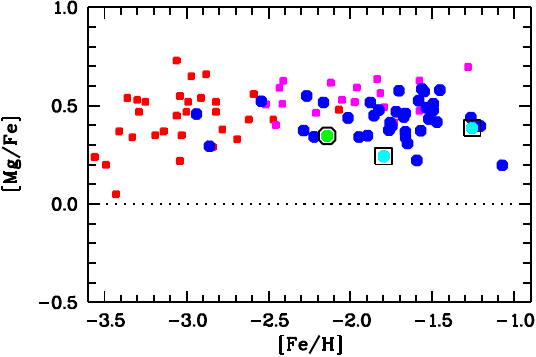}
\caption{[Mg/Fe] versus [Fe/H] for the observed sample (blue circles). The young star 
is shown as a green dot inside a black circle, and the PISN candidates are shown as light-blue dots inside black squares.
The comparison samples are the more metal-poor FS sample from \citep[][red squares]{cayrel04} 
and the sample of similar metallicity from \citet[][violet squares]{PristineXIX}.
}
\label{fig:mgfe}
\end{figure}

The stars in the sample are enhanced in the $\alpha$ elements, as expected in metal-poor stars 
(see Fig.\,\ref{fig:mgfe}). 
The corrections for departures from local thermodynamic equilibrium (NLTE) 
for Mg and Si in this sample of stars are not large \citep[see][]{mataspinto22,PristineXIX}.
The star RVS726 has a low [Si/Fe] ratio ($\rm [Si/Fe]=-0.21$), lower than the other $\alpha$ elements, which are still low.
The strong \ion{S}{i} lines of Mult.\,1 at 920\,nm fall in the REDU CCD, but the region is unfortunately contaminated by telluric absorption.
We then carefully inspected the analysis done by \mygi\ and rejected the problematic lines.
We then derived A(S) for nine stars. 
After applying the NLTE corrections on [S/H] from \citet{takeda05}, two stars were found to have a negative [S/Fe]:
$\rm [S/Fe]=-0.05$ for RVS703 (based on one single line) and $\rm [S/Fe]=-0.25$ for RVS712 (based on two lines with a relatively high scatter of 0.25\,dex).
We investigated several \ion{Ca}{i} lines to derive the Ca abundance.
The line-to-line scatter is generally small ($\rm \langle\sigma\rangle =0.10\pm 0.03$), as
expected from the S/N ratios of the spectra.
The NLTE corrections as provided by \citet{mashonkina16} are on the order of 0.1\,dex.
When deriving the NLTE corrections for the \ion{Ca}{i} for each star for lines used and available in \citet{mashonkina07}, 
we derived an average NLTE correction on A(Ca) of 0.10\,dex.

For the complete sample of stars for which we performed the chemical investigation (43 stars), we derived $\rm\langle[Ca/Fe]\rangle =0.34\pm 0.08$. 
For 41 of these stars, we derived an Mg abundance $\rm\langle[Mg/Fe]\rangle =0.42\pm 0.10$.
Removing the four stars with a consistently low Mg and Ca (RVS708, RVS709, RVS726 and RVS745), we derived 
$\rm\langle[Ca/Fe]\rangle =0.35\pm 0.06$ and $\rm\langle[Mg/Fe]\rangle =0.44\pm 0.08$, and the star-to-star scatter was reduced.
The star RVS708 also has low [Na/Fe] and [Al/Fe] ratios.
The star RVS709 has low [Na/Fe], [Si/Fe], [Y/Fe], and [Zn/Fe] ratios and a slightly low [Al/Fe].
The star RVS726 has low [Na/Fe] and [Si/Fe] ratios.

Titanium was derived from \ion{Ti}{ii} lines for all stars in the sample, while for all but one star (RVS725), we could derive A(Ti) from \ion{Ti}{i} lines.
In the stellar parameter of the stars we investigated, \ion{Ti}{ii} lines have small NLTE corrections, while the
NLTE corrections for \ion{Ti}{i} lines are non-negligible \citep[see][]{sitnova_ti,sitnova20}.
It was thus safer if we relied on \ion{Ti}{ii} lines for the Ti abundance.

We investigated the light elements (Na, Al). 
For Na we investigated the lines at 568.8, 589.0, 589.6, 616.0, 818.3, and 819.4\,nm, and we derived A(Na) for 38 stars in the sample.
We looked in the material provided by \citet{takeda03} to derive NLTE corrections.
For the sample, we obtained $\rm\langle[Na/Fe]\rangle =-0.13\pm 0.18$, with the lowest value being for RVS739 
($\rm [Na/Fe]=-0.57$, which becomes even lower when applying the NLTE correction on [Na/H]).
To investigate the Al abundance, we used the resonance lines at 394 and 396\,nm just for the star RVS738, while for the other stars
we used weak lines where the NLTE correction is not expected to be large \citep[see e.g.][]{baumueller97}.
The star RVS738 has $\rm [Al/Fe]=-1.07$ based on three lines and a line-to-line scatter below 0.02\,dex.

The single \ion{K}{i} line at 769.8\,nm that we investigated to derive K abundances is contaminated by telluric absorption in the case of stars
with large negative radial velocities. We thus decided to investigate this line through line profile fitting.
The other reason supporting this choice was that the NLTE correction is large for this line in the stellar parameter range
of the stars investigated here \citep{reggiani19}.
An investigation with \mygi\ fit the line with a model more metal rich by more than 0.5\,dex.
We derived the K abundance for 31 stars in the sample. 
We derived the NLTE corrections in \citet{reggiani19}, sometimes extrapolating in the tables provided.
The large [K/Fe] ratios ($\rm \langle[K/Fe]\rangle =+0.58\pm 0.17$) are due to large NLTE effects, but once they were taken into account
in [K/Fe], we obtained $\rm \langle[K/Fe]\rangle =+0.11\pm 0.14$.

In 41 stars of the sample, Mn was detected, and we found that all the stars have negative [Mn/Fe] ratios ($\rm\langle[Mn/Fe]\rangle =-0.28\pm 0.11$).
This is due to NLTE effects on the \ion{Mn}{i} lines \citep[see e.g.][]{bergemann08mn,bergemann19mn}.
We investigated the NLTE effects in \citet[][]{bergemann19mn} and applied them to the lines we used, and we found
corrections to always be positive, from about 0.1 to 0.43\,dex.
For all but one star, we could derive the Ni abundance, which is in excellent agreement with Fe
(for the 42 stars $\rm\langle[Ni/Fe]\rangle =-0.02\pm 0.05$).

We derived the Cu abundance for 21 stars in the sample, and they all have a negative [Cu/Fe] ratio ($\rm [Cu/Fe]=-0.52\pm 0.14$).
\citet{PristineXIX} investigated the NLTE effects on Cu for a sample of stars of similar parameters and derived an NLTE correction
on average slightly below +0.2\,dex, so when applying these NLTE corrections, the stars in our sample would still keep a negative[Cu/Fe] ratio.

\begin{figure}
\centering
\includegraphics[width=\hsize,clip=true]{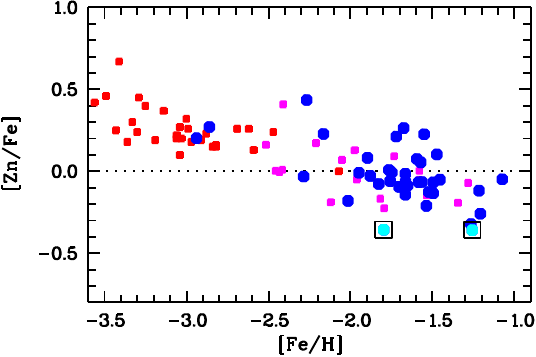}
\caption{[Zn/Fe] versus [Fe/H] for the observed sample.
Symbols are as in Fig.\,\ref{fig:mgfe}.
}
\label{fig:znfe}
\end{figure}

For 37 stars in the sample, Zn could be derived by investigating the \ion{Zn}{i} lines at 472.2 and 481.0\,nm (see Fig.\,\ref{fig:znfe}).
The NLTE corrections derived by interpolating in the values provided by \citet{sitnova_zn} are non-negative and small (not larger than 0.06\,dex).
When both lines are detected, the abundances derived from the two lines are generally in reasonable agreement.
Just for one star (RVS734), the two lines provide abundances at almost 0.5\,dex difference, but the S/N in this case is low (S/N=16 per pixel).
Three stars in the sample (RVS709, RVS714, and RVS727) show a $\rm [Zn/Fe]<-0.3$.
The star RVS709 is also poor in Cu ($\rm [Cu/Fe]=-0.82$; the lowest value in the sample), while the other two stars do not have any Cu detected.
These three stars are good candidates as pair-instability supernova (PISN) descendants \citep[see][]{salvadori19,PristineXIX,aguado23}.
The low Zn abundance could also point to the fact that these stars were accreted from a galaxy of such small mass to have allowed for 
a low enough star formation rate that the birth of massive stars was inhibited \citep[see][]{mucciarelli21}.

We derived the barium abundances by using line profile fitting. We used three \ion{Ba}{ii} lines:
585\,nm, 614\,nm, and 649\,nm. For the three lines, as recommended by \citet{heiter21}, we used
the oscillator strengths of \citet{davidson92}. Following \citet{korotin15}, we adopted
hyperfine splitting only for the 649\,nm line, and we adopted that in \citet{heiter21}. 
For all three lines, we interpolated the NLTE corrections in the table of \citet{korotin15}.
We were able to derive the barium abundance for 42 stars.
We investigated Europium through line profile fitting of the \ion{Eu}{ii} line at 381.9, 412.9, 664.5, and 742.6\,nm.
For 35 stars in the sample, we were able to derive A(Eu).
The [Eu/Fe] ratio (with Fe from \ion{Fe}{ii} lines) is always positive with $\rm\langle[Eu/Fe]\rangle =0.65\pm 0.21$.

We decided to use the two stars with differences in radial velocity from the two spectra in order 
to check the precision of this set of data and separately analyse the two spectra.
For RVS712, the average difference in the abundances we obtained from the two spectra 
is close to zero ($0.02\pm 0.06$\,dex), with the differences in the range $[-0.1,0.1]$\,dex.
Also for RVS736, the average difference we obtained by analysing the two spectra is close to zero ($0.02\pm 0.08$\,dex), with the differences in the range $[-0.1,0.15]$\,dex.
For Zn, whose abundance is based on the \ion{Zn}{i} line at 481.0\,nm, both spectra have an S/N around 15.
We concluded that the precision of the chemical investigation is between 0.1 and 0.2\,dex, according to the S/N of the spectra,
which are summarised in Table\,\ref{tab:snr}.

\subsection{Comparison with the RVS spectra}

We retrieved from the Gaia archive the RVS spectra for two stars in our sample: RVS719 and RVS733.
There are no astrophysical parameters for these two stars in Gaia\,DR3.
Both spectra have an S/N per pixel between 20 and 30 at 855\,nm, which are low for middle-resolution infra-red spectra.
We analysed the RVS spectra to derive Fe and Ti abundances.

For RVS719, from four \ion{Fe}{i} lines, we derived $\rm [Fe/H]=-1.26\pm 0.17$, to be compared to
$\rm [Fe/H]=-1.21\pm 0.22$ from the complete UVES spectrum. 
From one single \ion{Ti}{i} line, we derived $\rm [Ti/H]=-0.63$, to be compared to $\rm [Ti/H]=-0.70\pm 0.12$ from 25 lines
from the UVES spectrum. The \ion{Ti}{i} line investigated in the RVS spectrum was rejected in the UVES analysis because it is too strong, 
but it would provide [Ti/H]=-0.66.

For RVS733, we could only derive $\rm [Fe/H]=-1.32\pm 0.45$ from four \ion{Fe}{i} lines, to be compared to
$\rm [Fe/H]=-1.67\pm 0.19$ derived from the complete UVES spectrum. The agreement is well within the uncertainties.

\section{Discussion}

\begin{figure}
\centering
\includegraphics[width=\hsize,clip=true]{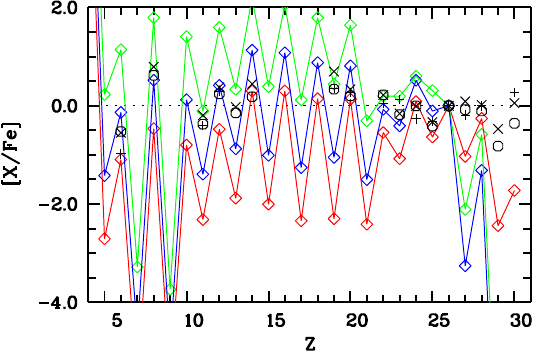}
\caption{[X/Fe] for abundances derived from neutral lines for a Zn-poor (RVS709, open circles), a Zn-normal (RVS728, cross symbols),
and a Zn-rich (RVS740, plus symbols) star.
{We also show the theoretical yields of \citet{HW2002}
for stellar masses of 168 $M_\odot$ (green lozenges and line),
195 $M_\odot$ (blue lozenges and line), and 242  
$M_\odot$ (red lozenges and line).}
}
\label{fig:elfeh}
\end{figure}

Three stars in the sample (RVS708, RVS709, RVS740) seem lower than the others in [Mg/Fe] and [Ca/Fe], while
RVS726 has a low [Ca/Fe] but no Mg detection, and RVS745 has low [Mg/Fe] and [Ca/Fe] ratios, but this is consistent with other stars.
This conjunction of low $\alpha$ elements and low [Zn/Fe] and [Cu/Fe] ratios makes RVS709 a perfect candidate 
as a PISN descendant.
\citet{salvadori19} investigated from a theoretical point of view the chemical characteristics
expected in a PISN descendant and highlighted that the effect is most evident at the metal-poor regime 
(with the most evident effect at $\rm -2\lesssim [Fe/H]\lesssim -1$) and 
overall on nitrogen, copper, and zinc.  
Since Cu and Zn abundance determinations are rare among chemical investigations in metal-poor stars, 
\citet{aguado23} built a way to use all the abundances to select stars that can be PISN descendants
and eventually observe them to derive A(Cu), A(Zn), and possibly also A(N).
Recently, \citet{xing23} discovered another PISN descendant in the Galaxy.
In Fig.\,\ref{fig:elfeh}, the [X/Fe] ratios (derived from neutral species) are compared for a Zn-poor star (RVS709), a Zn-normal star (RVS728),
and a Zn-rich star (RVS740).
The star RVS709 shows a systematic [X/Fe] ratio that is lower with respect to the other two stars.
In Fig.\,\ref{fig:elfeh}, we also show the theoretical yields
of \citet{HW2002} for three masses, 168, 195, and 242 $M_\odot$,
that end their lives as PISNs.
It is clear from this comparison that none of these stars may have
been formed from pure PISN ejecta.
If PISN ejecta are diluted with gas polluted by less
massive supernovae (SNe), then patterns like that of RVS709 can be
reproduced \citep[see][for an extensive discussion]{aguado2023}.

\begin{figure}
\centering
\includegraphics[width=\hsize,clip=true]{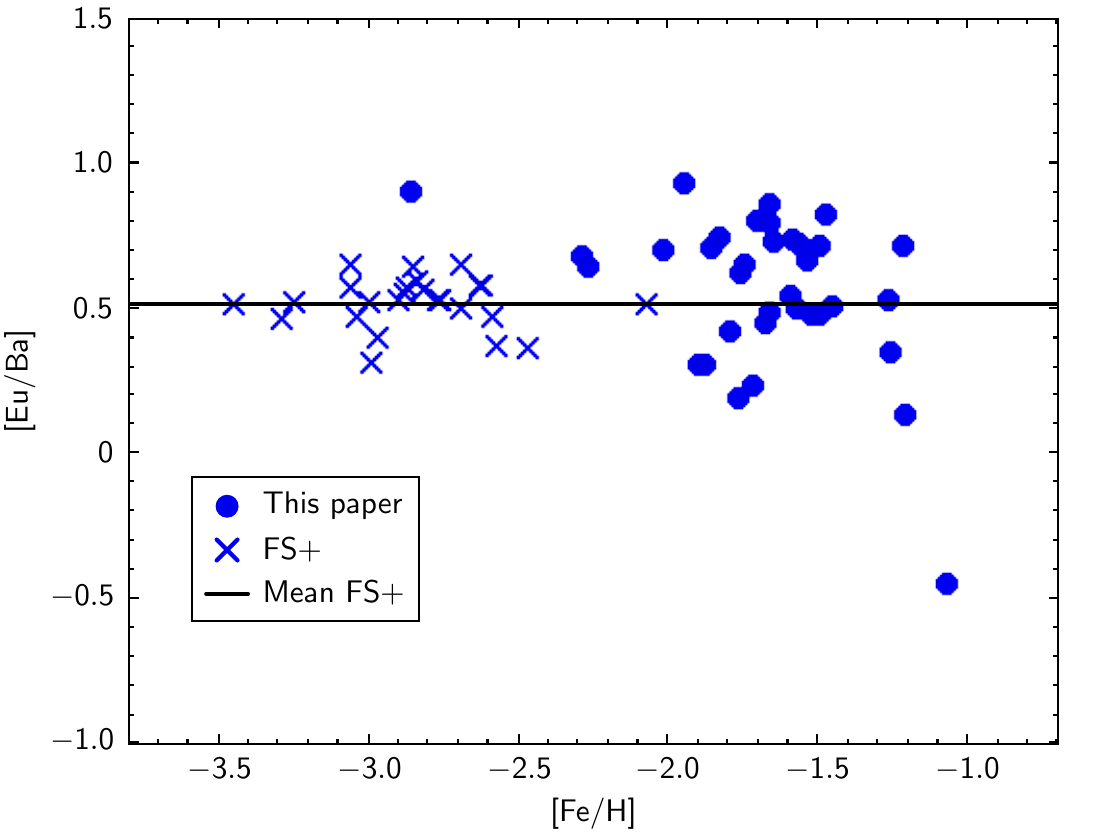}
\caption{[Eu/Ba] versus [Fe/H] diagram. For our stars, the NLTE corrections
to Ba have been applied. For 
comparison, we added stars from the First Stars sample (see text), 
labelled FS+.
The solid horizontal line corresponds to the mean
value of all the FS+ values.} 
\label{fig:EuBa}
\end{figure}

Europium is a pure $r-$process element,
while barium is mainly an $s-$process element \citep{arlandini99,prantzos20}. The ratio Eu/Ba
can be used as a diagnostic of the relative importance
of the two processes in any stellar population.
In Fig.\,\ref{fig:EuBa}, we show [Eu/Ba] as a function of [Fe/H]
for our programme stars as well as for the First Stars sample
\citep{FS5,FS7,FS16},
complemented with two measurements from \citet{sneden09} and \citet{hill17}, labelled
FS+ in the plot.
The FS+ stars show a constant [Eu/Ba] ratio of 0.52 with a tiny scatter
of 0.09\,dex. This strongly suggests that both Eu and Ba are 
synthesised in the same process. This value is not far from the
prediction of  \citet{wanajo07} of 0.62 for the hot $r-$process 
\citep[see also][Fig.\,15]{FS16}.
Of the five stars with metallicities above --1.3, three have [Eu/Ba]$< 0$.
We tentatively interpreted this as a sign of Ba $s-$process production
in AGB stars. For lower metallicity stars, however, the mean is very close to that
of the FS+ sample, albeit with a larger scatter that is compatible with the lower
S/N of our spectra with respect to those of the FS+ sample.
This does suggest that in the most metal-poor part of the sample ($\rm [Fe/H]<-1.3$), both Eu and Ba are formed only by the $r-$process,
and by inference, we may extend this hypothesis to all the neutron capture elements.

\citet{ghs1} and \citet{bonifacio23} investigated a sample of stars selected by their high speed with respect to the Sun.
They required in Gaia\,DR2 a transverse velocity higher than 500\,\kms.
The sample of stars we investigated was selected according to their high radial velocity
in order to verify their reliability (discussed in Katz et al. in preparation).
All of these stars \citep[of this paper and the stars discussed in][]{bonifacio23} are then expected to be halo stars.
All of these selections collected mainly metal-poor stars, with a paucity of extremely metal-poor stars,
and in the case of the stars analysed in this work, the most metal-poor star (RVS713) has $\rm [Fe/H]=-2.94$.
It is true that extremely metal-poor stars ($\rm [Fe/H]<-3$) are rare objects, but the samples containing the stars we investigated are quite large (about 400 stars), and just a handful of stars \citep[in][]{bonifacio23} have $\rm [Fe/H]<-3$.

Another similarity with the investigations of \citet{ghs1} and \citet{bonifacio23} is the presence of stars
compatible with a young age.
The two stars RVS725 and RVS726 seem to be younger than 1\,Ga, and they both are expected to have large masses.
As discussed in \citet{bonifacio23}, these stars can be evolved blue stragglers that resulted from the merger of three stars
(in the case of RVS725, perhaps even four).
The formation of a blue straggler by the merging of three stars is described by \citet{meyer80}, but as stated by \citet{fiorentino14},
massive blue stragglers are very rare objects.
To further support the small probability of forming a massive blue straggler from
a triple system, one can first consider the low fraction of triple systems
among solar-type stars ($9\% \pm 2\%$, \citealt{raghavan2010}) 
and the low probability that in a {\em destabilised} triple system,
a collision involving all three stars takes place \citep[0.6\%,][]{toonen2022}.
On the other hand, collisions involving only the inner pair are relatively common, up
to 24\%\ \citep{toonen2022}.
But these latter numbers are only for a destabilised triple system, which are only a fraction
of all triple systems. These considerations explain why blue stragglers
with masses larger than twice the turn-off mass are rare.
Thus, the possibility must be confronted that these stars are really young, formed in a recent accretion event at the moment of the infall
of the accreted object (a dwarf galaxy or a cluster) in the Milky Way.
If so, the age of these stars would provide us with the time of the accretion events.

The 43 stars we chemically investigated are all metal-poor ($\rm [Fe/H]<-1$), but none is a CEMP star.
This is surprising, as one would expect 21\% of the stars to be CEMPs at the very metal-poor regime, $\rm [Fe/H]<-2$ \citep{lucatello06}.
Clearly the numbers are small, but one would expect to see at least one star rich in carbon.

\section{Conclusion}

The chemical analysis of our sample of stars shows, by and large, the typical patterns
found in other samples of metal-poor stars.
What stands out is (i) there are up to three candidate PISN descendants with low $\alpha$ elements, Cu and Zn, and
(ii) two are apparently young stars that can hardly be interpreted as blue stragglers without
invoking the merging of multiple systems. 

We confirm the small number of extremely metal-poor stars among samples of high-velocity stars.
We point out that RVS721 lies within $10''$ of the track of the Gjoll stream \citep{ibata19} associated
with the globular cluster NGC\,3201.
The metallicity of the star is $-1.66\pm 0.16$\,dex, which is a good match
with the catalogue metallicity of the cluster, $-1.59$\,dex \citep{harris96}, thus supporting the association.  

\begin{acknowledgements}
The authors wish to thank Yoichi Takeda for providing useful material.
The authors wish to thank the referee.
We gratefully acknowledge support from the French National Research Agency (ANR) funded projects ``Pristine'' (ANR-18-CE31-0017)
and ``MOD4Gaia" (ANR-15- CE31-0007).
HGL gratefully acknowledges financial support by the Deutsche
Forschungsgemeinschaft (DFG, German Research Foundation)
Project ID 138713538-SFB 881 (``The Milky Way System'', subproject A04).
This work has made use of data from the European Space Agency (ESA) mission
{\it Gaia} (\url{https://www.cosmos.esa.int/gaia}), processed by the {\it Gaia}
Data Processing and Analysis Consortium (DPAC,
\url{https://www.cosmos.esa.int/web/gaia/dpac/consortium}). Funding for the DPAC
has been provided by national institutions, in particular the institutions
participating in the {\it Gaia} Multilateral Agreement.
This research has made use of the SIMBAD database, operated at CDS, Strasbourg, France.
\end{acknowledgements}





%


   \bibliographystyle{aa} 

   \bibliography{biblio} 

%


\begin{appendix}

%
\section{Variable stars}

\begin{figure}
\centering
\resizebox{0.9\hsize}{!}{\includegraphics{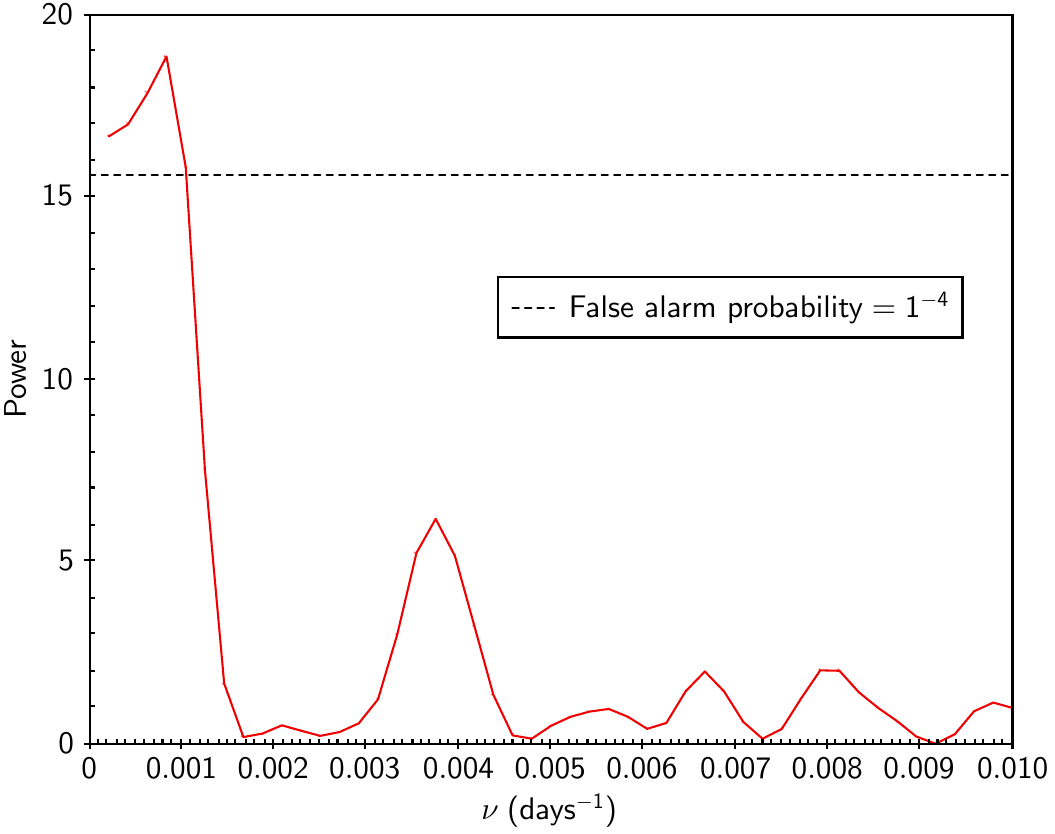}}
\caption{Lomb-Scargle periodogram of the $G$ epoch photometry  of RVS711.
The dotted line shows the level of false alarm probability $1.^{-4}$}
\label{LS_RVS711}
\end{figure}

Four stars are classified by Gaia DR3 as variables.
The stars RVS708, RVS719 and RVS724 are classified as long period variables,
with probabilities of over 70\%, while RVS711 is classified as an RS Canum Venaticorum (RS CVn) variable,
with a probability of 44\%.
No periods or other information beyond the epoch photometry is available from Gaia DR3.

\subsection{RVS711}
In our spectrum, the Ca{\sc II} H and K lines
show a clear emission core, supporting the classification as an RS CVn.
We computed the Lomb-Scargle periodogram using the $G$ data (see Fig.\,\ref{LS_RVS711}), and the only significative peak is evident
at 120.48 days, which is quite high for an RS CVn, since most of these stars
have periods below 20 days \citep{Martinez}. We tried to phase the data with this period
and were unable to obtain a smooth light curve. It is probable that the timeseries is too short
and sparsely sampled to provide an accurate period.
Under these conditions, it is impossible to understand in which phase our spectrum was
observed. However, the whole light curve spans a range of 0.05\,mag in $G_{BP}-G_{RP}$,
corresponding to less than 100\,K. This is an additional error to be considered
on our adopted effective temperature.

\subsection{RVS708}

This star has been classified as a long period variable
by \citet{2018AJ....156..241H}, named as  ATO J283.9396-17.3013, as well as by Gaia DR3.
\citet{2018AJ....156..241H} provided a period of 315.971905 days from a Lomb-Scargle periodogram.
We tried to phase the Gaia $G$ epoch photometry (24 points) with this period but failed 
to obtain a smooth light curve. The $G_{BP}-G_{RP}$ colour spans a range of 0.07\,mag
in the Gaia DR3 epoch photometry. This corresponds to about 150\,K and has to be considered
an additional error on our adopted effective temperature.
Our spectrum shows clear emissions in the \ion{Ca}{ii} H and K lines.

\subsection{RVS719}

For this star, there is no information in the literature.
We computed the Lomb-Scargle periodogram for the $G$ epoch photometry 
but could not identify any statistically significant peak.
The $G_{BP}-G_{RP}$ colour spans a range of about 0.09\,mag,
corresponding to almost 200\,K. This is an additional uncertainty on 
our adopted effective temperature.

\subsection{RVS724}

\begin{figure}
\centering
\resizebox{0.9\hsize}{!}{\includegraphics{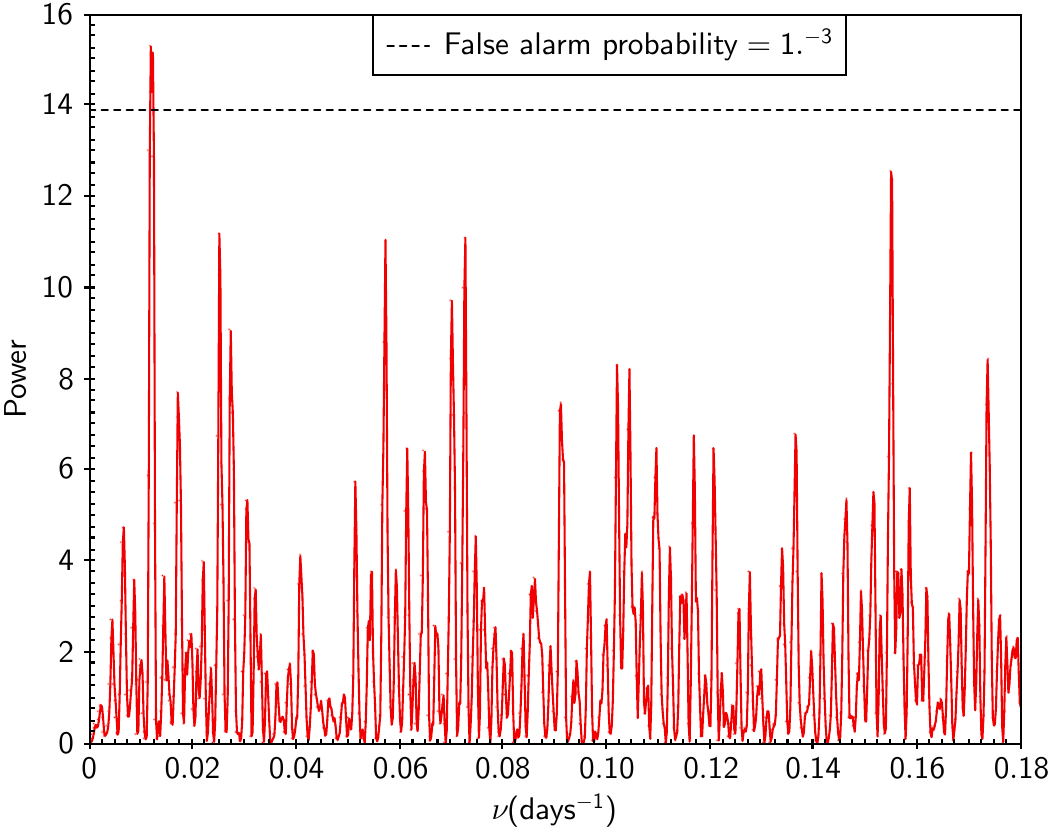}}
\caption{Lomb-Scargle periodogram of the $G$ epoch photometry of RVS724
The dotted line shows the level of false alarm probability $1.^{-3}$}
\label{LS_RVS724}
\end{figure}

\begin{figure}
\centering
\resizebox{0.9\hsize}{!}{\includegraphics{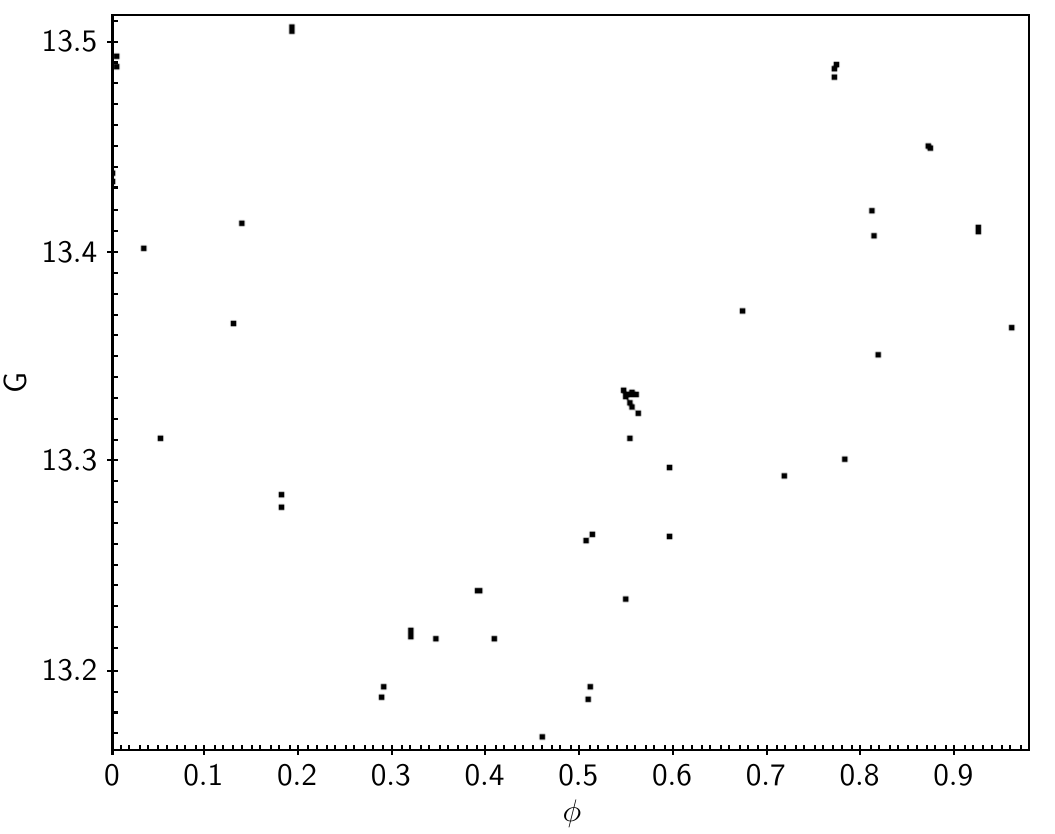}}
\caption{$G$ epoch photometry of RVS724 phased with a period  of 84.45946 days.}
\label{RVS724_phase}
\end{figure}

We computed the Lomb-Scargle periodogram using the $G$ epoch photometry,
and it is shown in Fig.\,\ref{LS_RVS724}.
There is only one peak higher than the false alarm probability
of $10^{-3}$, corresponding to a period of 84.45946 days.
We phased the $G$ epoch photometry with this period, and the result is
shown in Fig.\,\ref{RVS724_phase}. 
While the result is plausible, it is inconsistent with the precision
of the Gaia photometry. This suggests that a longer timeseries is required
to determine an accurate period, and we do not exclude that the star may be multi-periodic.
This discourages us to use this curve to estimate the phase at which we observed
the spectrum. In the epoch photometry, the $G_{BP}-G_{RP}$ colour spans a range slightly over 0.3 mag, which
corresponds to over 600\,K.
This implies that the effective temperature that we derived 
from the mean colours is very uncertain.
This is clear also from the two spectra at our disposal that provide different
abundances, very likely because they were observed at different phases and different effective
temperatures. Our  spectra display a considerable level of activity with
the presence of several emission lines.

\section{Details on some stars}

Some of the stars have a sufficient S/N in the blue wavelength region to allow for an emission to be seen
in the core of the \ion{Ca}{ii} H and K lines, namely, RVS700, RVS701, RVS703, RVS704, RVS705, RVS707, RVS708, RVS711, RVS719, RVS724, and RVS736.
In some stars, a P-cygni profile in $\rm H\alpha$ can be detected, including
RVS701, RVS703, RVS704, RVS707, RVS708, RVS709, RVS719, RVS724, RVS726, RVS732, RVS733, RVS740, and RVS743.
Some of the stars have already been investigated and are present in the literature.

\subsection{RVS701}

This star, as SMSS J120148.09-414126.6, has been investigated by \citet{dietz20},
who derived $\rm [Fe/H]=-1.196$. This is to be compared to $\rm [Fe/H]=-1.584\pm 0.178$ from our analysis.

\subsection{RVS705}

According to \citet{alksnis01}, this is a carbon star. The UVES spectrum confirms this finding.

\subsection{RVS707}

This star appears in the sample of \citet{dietz20}, with $\rm [Fe/H]=-1.708$. This is in excellent agreement with our
value of $\rm [Fe/H]=-1.758\pm 0.196$.

\subsection{RVS728}

This star is in the sample by \citet{dietz20}, with $\rm [Fe/H]=-1.460$. This is close to our value of $\rm [Fe/H]=-1.571\pm 0.171$.

\subsection{RVS733}

\citet{dietz20} derived for this star $\rm [Fe/H]=-1.496$, while \citet{hayes18} adopted (\teff/\logg/[Fe/H]) of 4392/1.0647/--1.340,
and \citet{jonsson20} adopted 4420/1.2902/--1.404. These values are to be compared to
our values of 4336/1.49/$-1.675\pm 0.194$.

\subsection{RVS743}

According to \citet{srinivasan16}, this is a red supergiant candidate of the Small Magellanic Cloud (SMC).
According to the Gaia\,DR3 parallax ($0.03654\pm 0.01038$), the star is too close to belong to the SMC.

\section{Abundances}

\begin{figure}
\centering
\includegraphics[width=\hsize,clip=true]{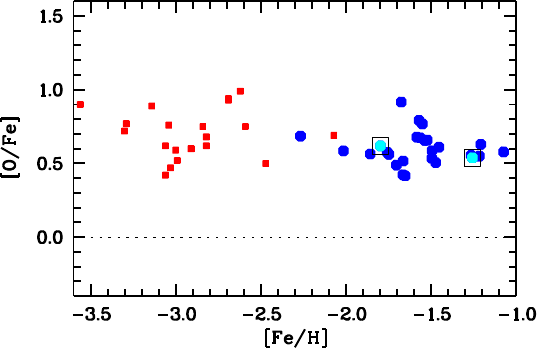}
\includegraphics[width=\hsize,clip=true]{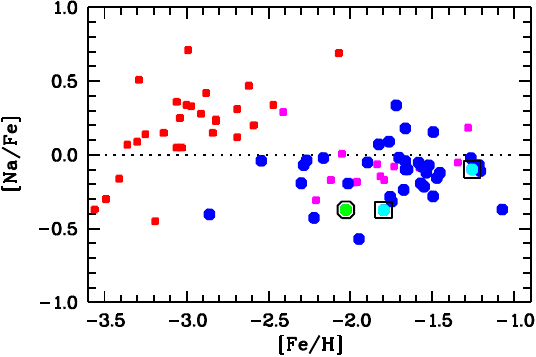}
\includegraphics[width=\hsize,clip=true]{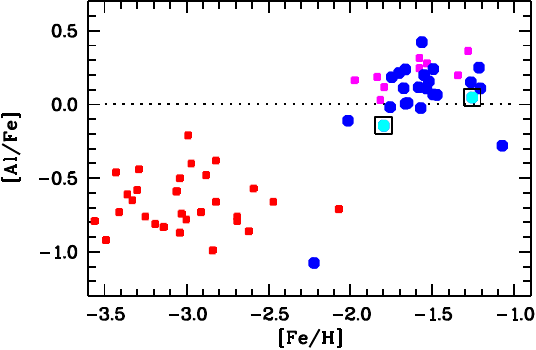}
\caption{[X/Fe] versus [Fe/H] (for O, Na and Al) for the observed sample (blue circles). The young stars 
are shown as green dots inside black circles, and the PISN candidates are shown as light-blue dots inside black squares.
The comparison samples are the more metal-poor FS sample from \citep[][red squares]{cayrel04} 
and the sample of similar metallicity from \citet[][violet squares]{PristineXIX}.
}
\label{fig:sife}
\end{figure}

The abundances are provided in an online table (Table\,C.1) available at CDS.
The table contains the name of the star, the Gaia\,DR3 identifier, the stellar parameters, the abundances, and the NLTE corrections discussed in the text.

The lines used in the star-by-star analysis for all elements are provided at CDS in an online table (Table\,C.2). 
We recall that \mygi\ performed a fit in chosen wavelength ranges. In the table, all the lines of the designated elements presented in the range are reported.
Except for the forbidden [OI] lines, we cut all the \ion{Fe}{i} lines at $\rm\log{gf}<-3.5$, and for the other elements, we cut them at $\rm\log{gf}<-4.5$.
We show the plots of [X/Fe] versus [Fe/H] for the abundances derived and provided in the online table 
compared to the sample from \citet{cayrel04} and \citet{PristineXIX} for the elements not already shown in the main text.

\begin{figure}
\centering
\includegraphics[width=\hsize,clip=true]{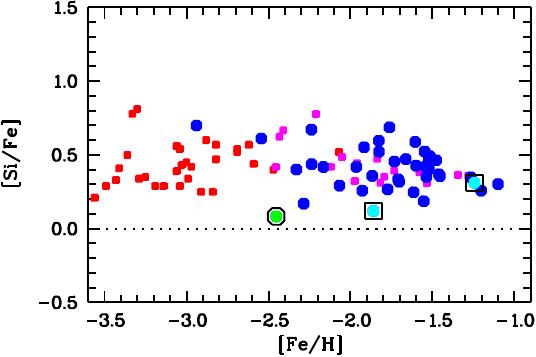}
\includegraphics[width=\hsize,clip=true]{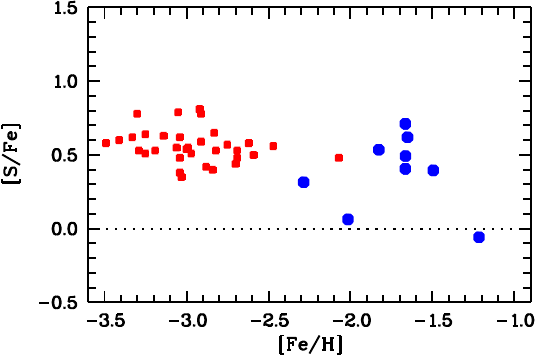}
\includegraphics[width=\hsize,clip=true]{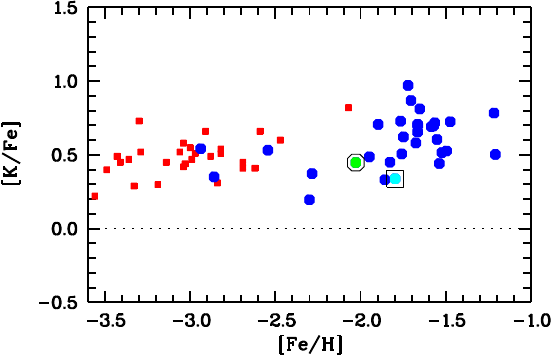}
\includegraphics[width=\hsize,clip=true]{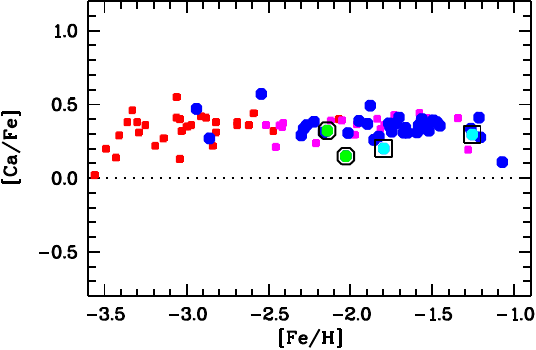}
\caption{[X/Fe] versus [Fe/H] (for Si, S, K, and Ca) for the observed sample.
Symbols and comparison samples are as in Fig.\,\ref{fig:sife}.
}
\label{fig:sfe}
\end{figure}

\begin{figure}
\centering
\includegraphics[width=\hsize,clip=true]{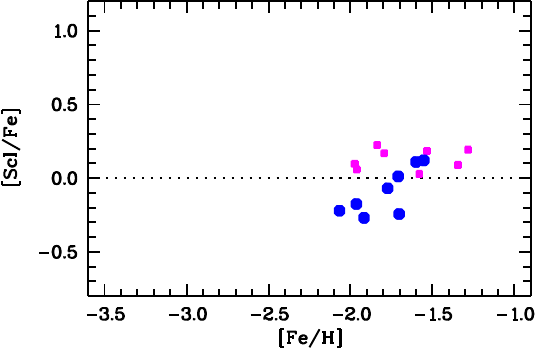}
\includegraphics[width=\hsize,clip=true]{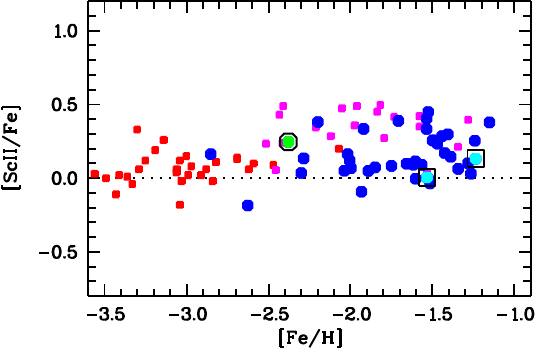}
\includegraphics[width=\hsize,clip=true]{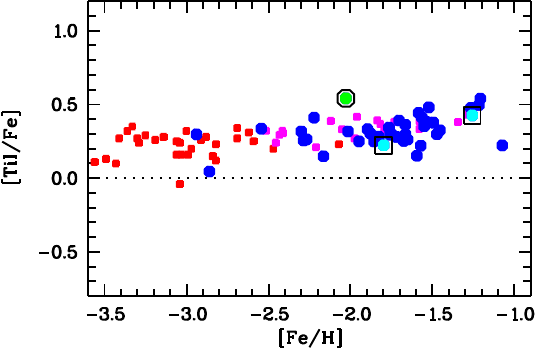}
\includegraphics[width=\hsize,clip=true]{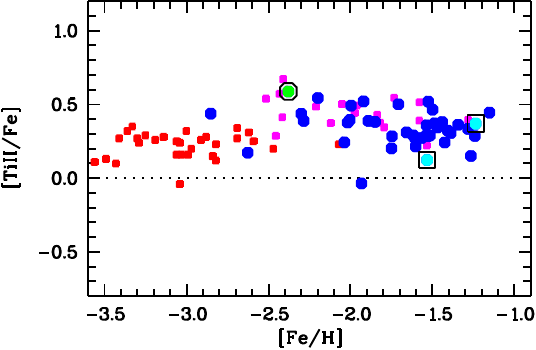}
\caption{[X/Fe] versus [Fe/H] (for Sc and Ti).
Symbols are as in Fig.\,\ref{fig:sife}. In [X/Fe] ratios for the ionised Sc and Ti, Fe is from \ion{Fe}{ii} lines.
}
\label{fig:tife}
\end{figure}

\begin{figure}
\centering
\includegraphics[width=\hsize,clip=true]{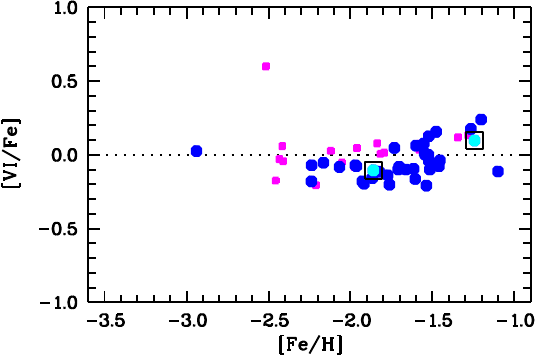}
\includegraphics[width=\hsize,clip=true]{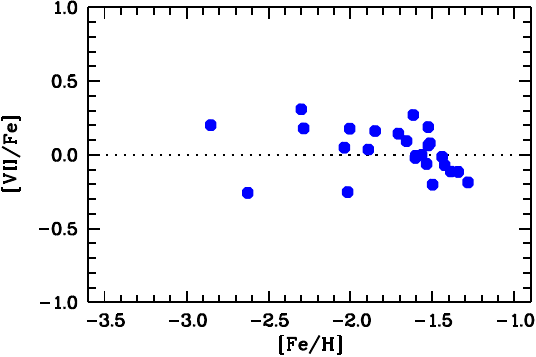}
\includegraphics[width=\hsize,clip=true]{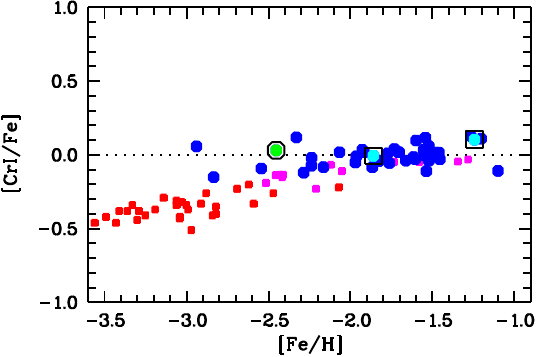}
\includegraphics[width=\hsize,clip=true]{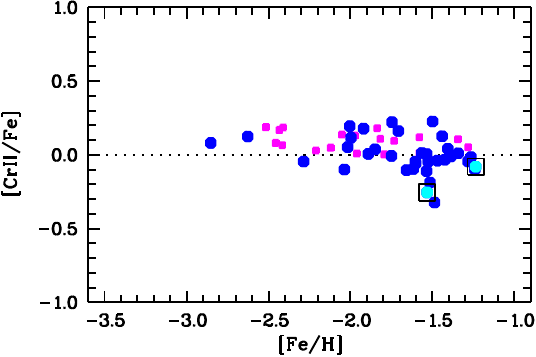}
\caption{[X/Fe] versus [Fe/H] (for V and Cr).
Symbols are as in Fig.\,\ref{fig:sife}. In [X/Fe] ratios for the ionised V and Cr, Fe is from \ion{Fe}{ii} lines.
}
\label{fig:vfe}
\end{figure}

\begin{figure}
\centering
\includegraphics[width=\hsize,clip=true]{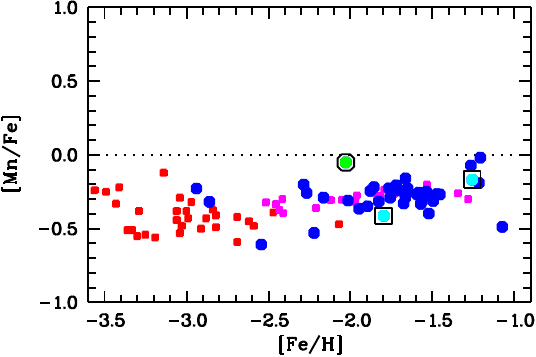}
\includegraphics[width=\hsize,clip=true]{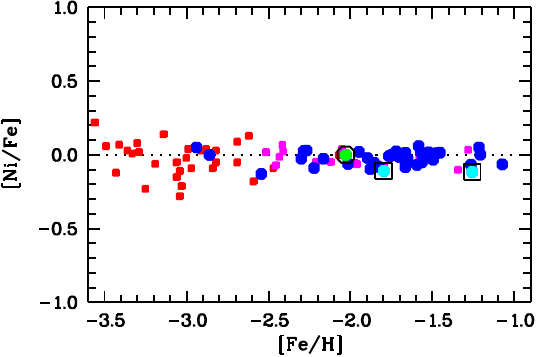}
\includegraphics[width=\hsize,clip=true]{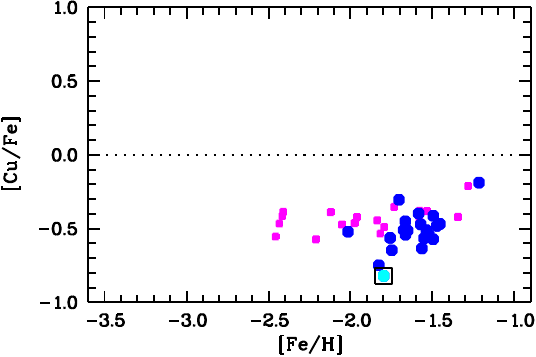}
\includegraphics[width=\hsize,clip=true]{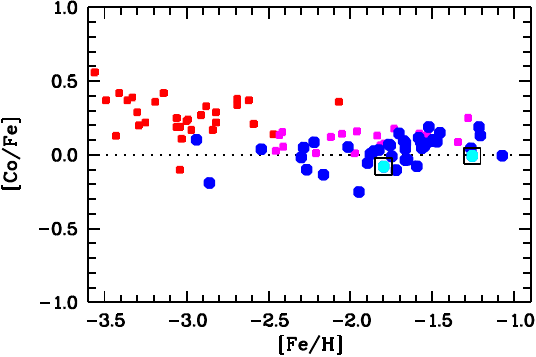}
\caption{[X/Fe] versus [Fe/H] (for Mn, Ni, Cu and Co).
Symbols are as in Fig.\,\ref{fig:sife}. 
}
\label{fig:mnfe}
\end{figure}

\begin{figure}
\centering
\includegraphics[width=\hsize,clip=true]{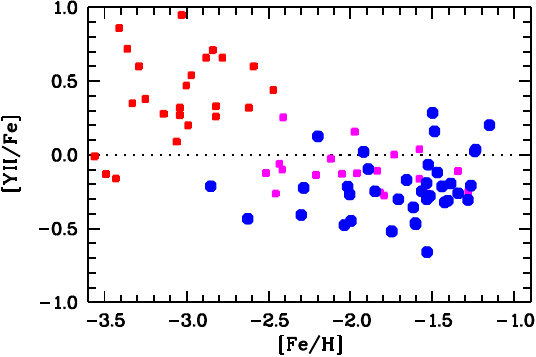}
\includegraphics[width=\hsize,clip=true]{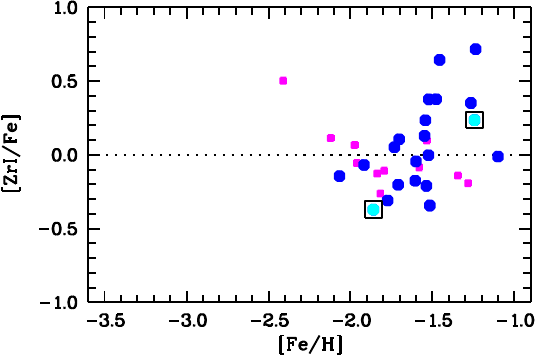}
\includegraphics[width=\hsize,clip=true]{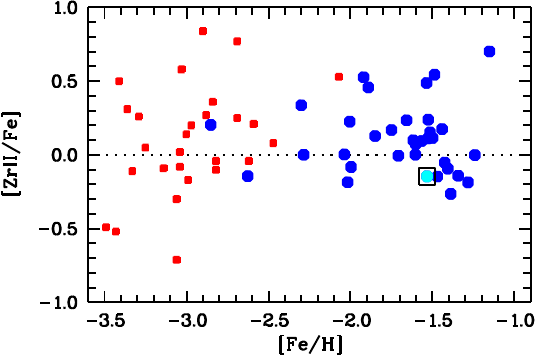}
\caption{[X/Fe] versus [Fe/H] (for Y and Zr).
Symbols are as in Fig.\,\ref{fig:sife} except that here the FS sample is from \citet{FS7}. 
In [X/Fe] ratios for the ionised Y and Zr, Fe is from \ion{Fe}{ii} lines.
}
\label{fig:yfe}
\end{figure}

\end{appendix}
\end{document}